\documentclass{aa}  
\usepackage{natbib}

\usepackage{graphicx}
\usepackage{txfonts}
\newcommand{\logg}{\ensuremath{\log g}}

\begin{document} 

   \title{New ultra metal-poor stars from SDSS:  
follow-up GTC medium-resolution spectroscopy
\thanks{
 Based on observations made with the Gran Telescopio Canarias (GTC), 
 installed in the Spanish Observatorio del Roque de los Muchachos 
 of the Instituto de Astrof\'{\i}sica de Canarias, on the island of La Palma.
 Programme ID GTC2E-16A and ID GTC65-16B.}
 }


      \author{D.~S. Aguado\inst{1,2}, C. Allende Prieto\inst{1,2}, J.~I. Gonz\'alez Hern\'andez\inst{1,2}, R. Rebolo\inst{1,2,3}, E. Caffau\inst{4} }
   \institute{Instituto de Astrof\'{\i}sica de Canarias,
              V\'{\i}a L\'actea, 38205 La Laguna, Tenerife, Spain\\              
         \and
             Universidad de La Laguna, Departamento de Astrof\'{\i}sica, 
             38206 La Laguna, Tenerife, Spain \\  
         \and
             Consejo Superior de Investigaciones Cient\'{\i}ficas, 28006 Madrid, Spain\\
              \and
GEPI, Observatoire de Paris, PSL Research Univsersity, CNRS, Place Jules Janssen, 92190 Meudon, France\\
            }   
  
\authorrunning{D.S. Aguado et al.}
\titlerunning{New ultra metal-poor stars observed with GTC}

 
  \abstract
   {The first generation of stars formed in the Galaxy left behind the chemical signatures of their nucleosynthesis in the interstellar
   medium, visible today in the atmospheres of low-mass stars that formed afterwards. Sampling the chemistry of those low-mass provides insight into the first stars.}
   {We aim to increase the samples of stars with extremely low metal abundances, 
   identifying ultra metal-poor stars from spectra with modest spectral resolution and signal-to-noise ratio (S/N).
   Achieving this goal involves deriving reliable metallicities and carbon abundances from such spectra.}
   {We carry out follow-up observations of faint, V$>$19, metal-poor candidates selected from SDSS 
   spectroscopy and observed with the Optical System for Imaging and low-Intermediate-Resolution Integrated Spectroscopy (OSIRIS) at GTC. The SDSS and follow-up OSIRIS spectra were analyzed 
   using the FERRE code to derive effective temperatures, surface gravities, metallicities and carbon abundances. 
   In addition, a well-known extremely metal-poor star has been included in our sample to calibrate the analysis
   methodology.}
   {We observed and analyzed five metal-poor candidates from modest-quality SDSS spectra. All stars
   in our sample have been confirmed as extremely metal-poor stars, in the  $\left[{\rm Fe/H}\right]<-3.3$ regime.
   We report  the recognition of J173403+644632, a carbon-enhanced ultra metal-poor dwarf  star with $\left[{\rm Fe/H}\right]=-4.3$ and $\left[{\rm C/Fe}\right]=+3.1$.}
   
   {}

\keywords{stars: Population II – stars: abundances – stars: Population III – 
Galaxy: abundances – Galaxy: formation – Galaxy: halo
               }

   \maketitle
%

\section{Introduction}\label{intro}
Metal-poor stars are extremely rare objects. Their study is key to understanding the early Galaxy. The fraction of stars with low metal content increases with magnitude (Robin et al. 2003) and increases at high Galactic latitudes. In addition, the number of halo stars at [Fe/H]$\sim -4$ is only a small percent of those 
at [Fe/H]$\sim -3$, and those are only a small percent of those at [Fe/H]$\sim -2$ \citep{alle16}.
About two dozen stars are known at [Fe/H]$<-4$ \citep{boni15,pla15,agu16} even though it is the most 
interesting regime on the yields from the very first generation of stars. Developing new procedures to identify faint metal-poor stars is an important task.

We have been trying to identify these rare objects from large ongoing spectroscopy surveys
(SDSS, LAMOST, etc.), to reobserve them at medium-resolution, and to derive reliable stellar
atmospheric parameters, including metallicity and whenever possible carbon abundances \citep{agu16,agu17I}.
This method can be used to identify extremely metal-poor stars as faint as V$\sim$19. 
Before performing high-resolution spectroscopy to derive multiple chemical abundances, having an intermediate medium-resolution step provides a powerful time-saving tool with regard to large telescopes (6-10\,m class)
 since it dramatically removes false positives in the analysis of 
large-scale spectroscopic surveys.
With this methodology, a large sample of ultra-metal poor stars can be assembled in preparation for the
arrival of the next 30\,m class generation of telescopes. 

This is the third paper in a series devoted to follow-up spectroscopy of extremely and ultra metal-poor stars 
identified from SDSS and other large-scale spectroscopic surveys. In this work we report on faint candidates 
observed with the Optical System for Imaging and low-Intermediate-Resolution Integrated Spectroscopy 
(OSIRIS) at Gran Telescopio de Canarias (GTC). We also report on the recognition of J173403+644632, which at g$_{mag}$=19.6 is
the faintest carbon-enhanced ultra metal-poor dwarf star known. The target selection is explained in Section \ref{selection}, which provides
a full description of the procedure originally described in \cite{agu16}.
The observations and data reduction are detailed in Section \ref{obs}.
The derived atmospheric parameters are analyzed and discussed in Section \ref{anal}.
In addition, a study of the most metal-poor star known, J1029+1729, is developed to test our methodology 
in Section \ref{caffau}.
Conclusions are given in Section \ref{discuss}.

\begin{table*}
\begin{center}
 \renewcommand{\tabcolsep}{5pt}
\centering

\caption{Coordinates and atmospheric parameters for the program 
stars based in the analysis of the low-resolution spectra with the FERRE code. The SDSS identification and parameters derived by the automatic SSPP are listed below (when available).}
\label{basic}
\begin{tabular}{lccccccccccc}
\hline
Star & $g$ &      RA &      DEC &$T_{\rm eff}$ & $\log g$  & $\left[{\rm Fe/H}\right]$ 
& $\left[{\rm C/Fe}\right]$ &$\rm <S/N>^{a}$\\
     & mag & h  '   ''&$\mathring{}$  '  ''   & K & $\rm cm\, s^{-2}$ &               &                       \\
\hline\hline
SDSS  J012512+070319   & 19.0 & 01:25:12.5 & +07:03:19.8 &6454 &4.9&-3.2 &0.8 & 22 \\
SDSS  J015131+163944   & 18.9& 01:51:31.2 & +16:39:44.9&  5917 &3.9&-3.6 &1.3   &19 \\
SDSS  J041800+062308   & 19.1& 04:18:00.7  &+06:23:08.1 &5579 &0.5& -4.3 & $-$& 21 \\
SDSS  J094708+461010   & 19.1 & 09:47:08.2& +46:10:10.1 & 5916&4.9&-4.7 & 0.2& 14 \\
SDSS  J173403+644632   & 19.6& 17:34:03.9& +64:46:32.9 &6470 &4.9&-3.8 & -0.2 &  15\\

\hline\hline
Star & plate &     mjd &      fiberid &$T_{SSPP}$& $\log g_{SSPP}$ &$\left[{\rm Fe/H}\right]_{SSPP}$ \\
&&&&K& $\rm cm\, s^{-2}$&\\
\hline\hline
SDSS  J012512+070319   & 4556&	55912&	180&$-$&$-$&$-$\\
SDSS  J015131+163944 & 5118& 55830 &336&$-$&$-$&$-$  \\
SDSS  J041800+062308  & 2826  &54389 &171&6387&2.70&-3.26\\
SDSS  J094708+461010 &4695&55957&944&$-$&$-$&$-$  & \\
SDSS  J173403+644632 &2561&54597& 501&6229&3.11 &-3.02 &\\

\end{tabular}
\end{center}

 \tablefoot{ $^a$ Signal-no-noise ratios have been calculated as the average value for the entire SDSS spectrum}

\end{table*}

\section{SDSS analysis and target selection}\label{selection}

We have re-analyzed more than a million spectra ($\sim 90\%$ stars)  from the original Sloan Digital 
Sky Survey \citep{yor00}, the Sloan Extension for Galactic Understanding and Exploration (SEGUE,
\citealt{yan09}) and the Baryonic Oscillations Spectroscopic Survey (BOSS, \citealt{eis11,daw13}).
The main stellar parameters (effective temperature $T_{\rm eff}$, surface gravity 
$\log g$, and metallicity [Fe/H]\footnote{We use the bracket notation to report
chemical abundances: [a/b]$ = \log \left( \frac{\rm N(a)}{\rm N(b)}\right) - \log
\left( \frac{\rm N(a)}{\rm N(b)}\right)_{\odot}$,
where $\rm N$(x) represents number density of nuclei of element x.}) 
and the carbon abundance have been derived.

The SDSS optical spectra are from SDSS Data Release 9 (DR9, \citealt{dr9})
for observations with the original SDSS spectrograph, 
and DR12 \citep{dr12} for data obtained with the upgraded BOSS spectrographs \citep{smee13}.
The resolving power of SDSS spectra is about 2,000, covering the range $\sim$3800 -- 9100\,\AA\,. 
Since the integration time is the same for all the targets, the S/N is widely variable.
We use the \emph{fortran} FERRE  code \citep{alle06} to perform a massive analysis of the spectra, as explained
in \citet{agu16} and \citet{agu17I}. For this work we focus on promising candidates with low S/N 
spectra and, consequently, higher error bars in the parameter determination  (See Table \ref{basic}). These targets are too faint for 
follow-up with the 4.2\,m WHT, and so a parallel program was started on the 10.4\,m GTC.

\section{Observations and data reduction}\label{obs}

We performed medium-resolution, long-slit spectroscopy with OSIRIS at the 10.4\,m telescope GTC. 
The program was granted 28 hours under proposals ID GTC2E-16A and ID GTC65-16B in service mode 
to carry out long-slit spectroscopy with the R2500U grism and a 1.0 arcsecond slit,  providing a spectral
range 3600-4500\,\AA\ with a resolving power (R$\equiv \lambda/\delta\lambda \sim$2300).
The slit was in all cases oriented at the paralactic angle.
The individual exposures were never longer than 30 minutes to
minimize the impact of cosmic rays (See Table \ref{observations}).

Data reduction consisted in bias substraction, flat-fielding, wavelength 
calibration using HgAr $+$ Xe calibration lamps, 
and a combination of individual spectra. The processing was performed with the \emph{twodspec} and \emph{onespec} 
packages in IRAF\footnote{IRAF is distributed by the National Optical Astronomy Observatory, 
which is operated by the Association of Universities for Research in Astronomy 
(AURA) under cooperative agreement with the National Science Foundation.} \citep{tod93}.
Cross-correlation between individual spectra and correction of radial velocity offsets was made
with the \emph{rv} package.

\begin{table}
\caption{Observing log.} 
\label{observations}      
\centering 
  
\begin{tabular}{c c c c c c c c}        
\hline\hline               
  Stars                      & Date  &    $\rm N_{\rm exp}$ & $\rm t_{\rm exp}$     &   seeing   \\ 
 &&&s\\
 \hline                                                                                 
    J012512+070319             & 27-Sep-2016 &  6        &  1600   &0.8 \\
                                                                                        
     J015131+163944           & 22/28-Sep-2016  &   7     &   1600  & 1.1 \\
                                                                                        
  J041800+062308  & 26/28-Sep-2016  & 6       &  1600   & 1.0\\
                                                                                        
    J094708+461010          & 28/31-Mar-2016  &   10      & 1827   & 0.9 \\
                                                                                        
  J173403+644632  & 29/30-Apr-2016  &  10       & 1827   & 1.0 \\
 \hline
 G64-12  & 1-Jan-2017  &    3     & 30    & 1.2\\
\hline 

\end{tabular}
\tablefoot{A slit width of 1 arcsec}\\
The moon was gray during all the observations and clouds  were not noticeable.

\end{table}


\section{Analysis and discussion}\label{anal}
\begin{table*}
\begin{center}
\renewcommand{\tabcolsep}{5pt}
\centering
\caption{The stellar parameters and main results obtained from OSIRIS spectra.
\label{AnalysisResults}}
\begin{tabular}{lcccccccccccc}
\hline
Star            & $\rm \langle S/N\rangle$ & ${\rm T_{\rm eff}}$ &$\Delta\rm T_{\rm eff}$  &\logg &$\Delta \logg$  &  $[\rm Fe/H]$ & $\Delta[\rm Fe/H] $ &$[\rm C/Fe]$ &$\Delta[\rm C/Fe] $\\
                     &  & [K] & [K]& [cm\, s$^{-2}$] & [cm\, s$^{-2}$]& & \\

\hline\hline
SDSS J012512+070319  &77  &6312&34   &5.0&0.07 &-3.3&0.1 & 1.2&0.6\\  
SDSS J015131+163944$^{(1)}$ &62  &6036&26  &5.0&0.07 &-3.8&0.08 & 1.1&0.5\\  
SDSS J041800+062308  &73  &6247&35  &5.0&0.08 &-3.4&0.1 & 0.7&0.7\\  
SDSS J094708+461010  &80 &5858&19   &5.0&0.07 &-4.1&0.06 & 1.0&0.3\\  
SDSS J173403+644632  &60 &6183&35   &5.0&0.09 &-4.3&0.11 & 3.1&0.1 \\ 

\hline
\hline              
G64$-$12            &132&6393&29&4.8&0.07&  -3.2& 0.1&  1.0&  0.4   \\
\hline              
\end{tabular}
\end{center}
\tablefoot{$\Delta$ are the internal uncertainties of  parameters derived with FERRE.}
\tablefoot{$(1)$ \citet{agu17I}}
\end{table*}

Following the same procedure described by \citet{agu17I}, we employed a grid of synthetic spectra 
computed with the ASS$\epsilon$T code \citep{koe08}. The model atmospheres were computed as in \citet{mez12}.
The spectral synthesis was performed as described by \citet{alle14}, but the grid we used differs
in the parameter coverage: $-6\leq[\rm Fe/H]\leq-2$; $+1\leq[\rm C/Fe]\leq+5$; $4750\,\rm K\leq T_{eff}\leq7000\,\rm K$, and
$1.0\leq \logg \leq5.0$. The $\alpha$-element abundance and micro-turbulence were fixed at $[\alpha/\rm Fe]$=0.4
and 2\,km s$^{-1}$, respectively. Lacking a reliable flux calibration for our observations, a 
running-mean filter with a window of 30 nearby pixels was used for continuum normalization. 
The same algorithm was applied to both observed 
and synthetic spectra after resampling the latter to the wavelength OSIRIS spectra of the former. 
Further details are given in \citet{agu17I}.
FERRE is used to simultaneously derive  the three main stellar parameters --effective temperature, surface gravity, 
metallicity-- as well as the carbon abundance.

Table \ref{AnalysisResults} summarizes the results of the FERRE analysis for the sample stars. 
A high-quality OSIRIS spectrum of the well-known metal-poor star G64-12 was analyzed as a reference.  The good agreement between the optimal 
set of parameters ($T_{\rm eff}= 6393$ K, $\log g=4.8$, [Fe/H]=$-3.2$) 
and the literature \citep{aoki06I,pla16} ($T_{\rm eff}= 6390$ K, $\log g=4.38$, [Fe/H]=$-3.2$), together 
with the quality of the fit, suggest that this is a highly reliable method of confirming extremely metal-poor stars. 
Further details on FERRE tests are given by \citet{alle14,agu17I}.

FERRE offers multiple options for deriving internal uncertainties. For our program we chose the estimates
obtained from Monte Carlo experiments after injecting noise into the observations and the error treatment is the same as that included in \citet{agu17I}. We derived the error bars by adding in quadrature the internal uncertainties showed in Table \ref{AnalysisResults} with  other sources of error:
$\Delta\rm T_{\rm eff}$=70\,K, $\Delta \logg$=0.5,  $\Delta[\rm Fe/H]$=0.1\,dex  and $\Delta[\rm C/Fe]$=0.2\,dex. 
The spectral range in common between the ISIS and the OSIRIS  spectra is about 820\,\AA\ wider, with the $\rm H_{\beta}$ 
line missing in the latter case. 
We examine  the results for each object in Section \ref{star}.

\subsection{Testing the methodology with J1019+1729}\label{caffau}
\begin{figure}
\begin{center}
{\includegraphics[width=105 mm, angle=180 ]{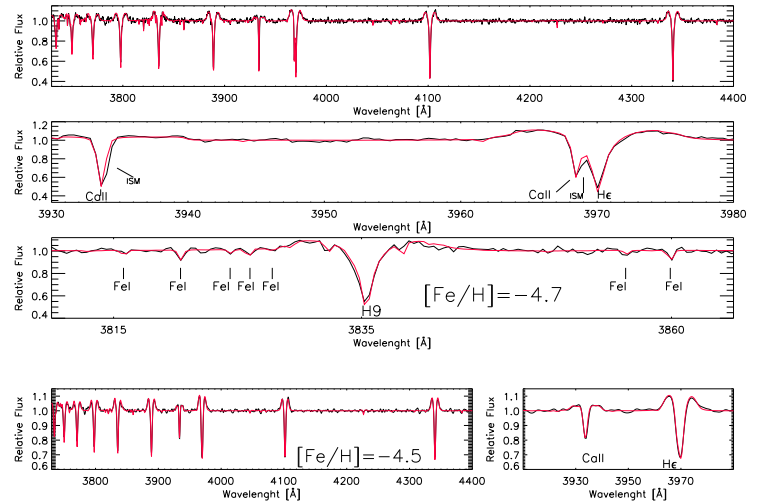}}
%
\end{center}
\caption{X-SHOOTER/VLT spectrum ($R=7900$) of J1019+1729 (black line) and the best fit derived by FERRE (red line). 
\emph{From top to bottom}: the entire spectrum  (3700\,\AA-4400\,\AA). The range of the \ion{Ca} {II} lines. 
The H9 spectral region with several \ion{Fe} {I} lines assuming $\rm T_{eff}=5834$\,K, $\logg=5.0$ 
from our previous analysis over the entire spectra. The derived iron abundance is displayed.
Finally the X-SHOOTER spectrum smoothed to the OSIRIS resolution ($R=2500$) and re-analyzed by FERRE: 
left panel shows the entire spectrum together with the derived metallicity while in the right panel 
a detail of the \ion{Ca} {II} lines are depicted.}
\label{caff}
\end{figure}

\citet{caff11,caff12I} recognized J1019+1729, a dwarf star at a metallicity of about [Fe/H]$\sim -5$, 
from SDSS/SEGUE  
spectroscopic data. The low carbon and oxygen abundances of this star challenge the picture of star formation that suggests no low-mass stars
were formed \citep{bro03}.
To test or methodology, we analyzed the X-SHOOTER 
spectrum of this star ($R=7900$ in the UVB arm). FERRE derives $\rm T_{eff}=5834$\,K,  $\logg=5.0$,  [C/Fe]$=-0.47$, and
[Fe/H]$=-4.55$. The effective temperature derived by the authors (5811\,K)
is perfectly recovered by the code and is in agreement with our analysis of the ISIS spectrum of this target presented in \citet{agu17I}.
In addition, FERRE is able to identify J1029+1729 as a dwarf star. 
Since it is not visible, we are able to place an upper limit [C/Fe]$<+0.7$.
However, our derived metallicity is higher than that determined by \citet{caff12I}, due to the ISM contribution to the \ion{Ca} {II} H and K lines.
The metallicity is compatible with our analysis in \citet{agu17I}
with ISIS and UVES-smoothed data. 
Furthermore, we ran the FERRE code only in the spectral region where the 
strongest Fe I lines are (3810\,\AA-3865\,\AA) and assuming
the previously derived $\rm T_{eff}$ and  $\logg=5.0$ over the entire X-SHOOTER spectrum, and 
we were able to recover [Fe/H]$=-4.75$ (See Fig. \ref{caff}), which is in agreement with the 
1D-LTE Fe I abundance derived by the authors in the original paper from higher resolution UVES spectroscopy.
Finally, we smoothed the X-SHOOTER spectrum of J1019+1729 to the OSIRIS resolution ($R=2500$) 
and, again, we ran FERRE and arrived at the values [Fe/H]$=-4.5$ and $\rm T_{eff}=5826\,K$.
This exercise demonstrates that our methodology is robust and works even at higher resolution ($R=7900$). 
We are able to recover the same metallicity and effective temperature from the analysis of 
SDSS, ISIS, and UVES-smoothed data \citep{agu17I}, and from X-SHOOTER and X-SHOOTER-smoothed spectra.

\subsection{Discussion of the new set of faint metal-poor stars}\label{star}
\begin{figure*}
\begin{center}
{\includegraphics[width=170 mm]{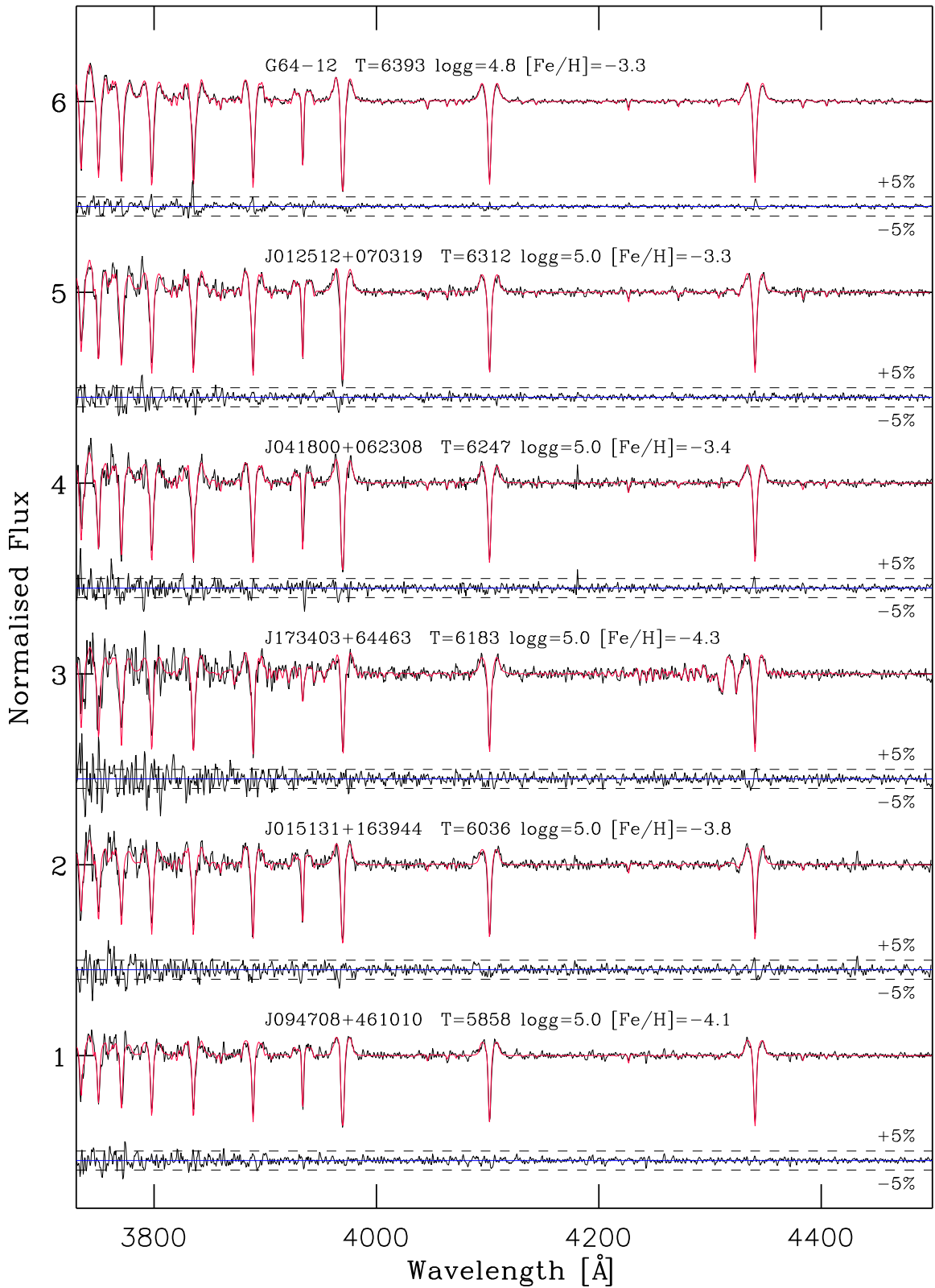}}
\end{center}
\caption{OSIRIS/GTC spectra (3700\,\AA-4500\,\AA) our stellar sample
(black line) and the best fit calculated with FERRE (red line). Under each of the spectra we also depict 
the residuals (difference between the observed spectrum and the best fit) together with the 
$+5\%$ and $-5\%$ reference lines. On top of each spectrum the main stellar parameters
are displayed.}
\label{all}
\end{figure*}

Figure \ref{all} (upper panel) shows the spectrum of G64-12 
(black solid line) and its best fit derived with FERRE. All spectra analyzed in this work are plotted below.

\begin{figure*}
\begin{center}
{\includegraphics[width=160 mm]{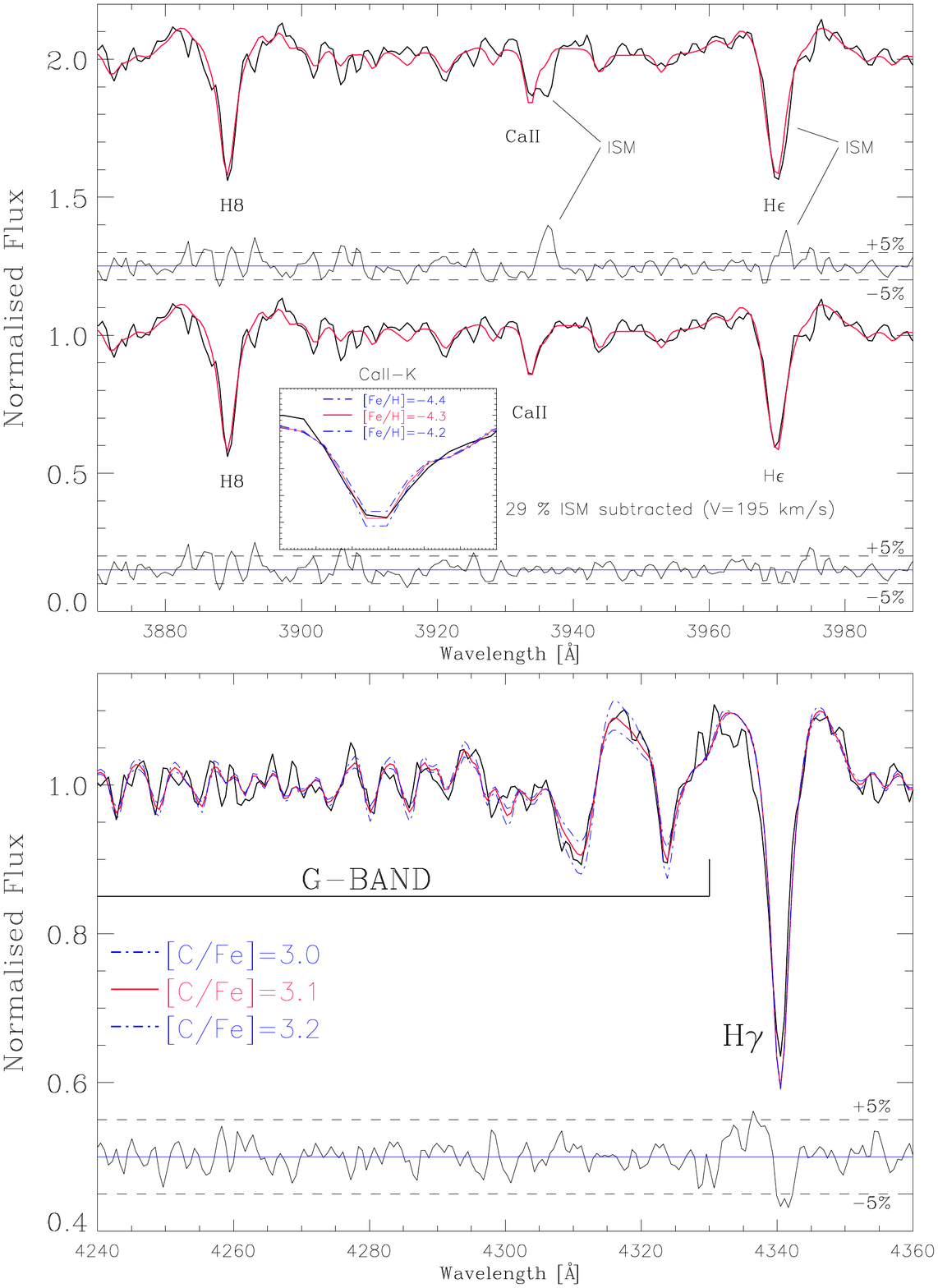}}
\end{center}
\caption{Upper panel: spectral range in the vicinity of the \ion{Ca} {II} H-K lines for the OSIRIS spectrum of the star 
J134157+513534 (black solid line) with the best fit (red line) before ISM subtraction (top panel) and after 
ISM subtraction (bottom panel). 
Below the spectra we display the residuals  together with the $+5\%$ and $-5\%$ reference lines.
A detail of the \ion{Ca} {II} K line after ISM subtraction with an upper and lower metallicity limit 
(dashed blue line) is plotted. Lower panel: spectral range of the G-band of the OSIRIS 
spectrum of J134157+513534 (black solid line) with the synthetic spectra for the best fit (red line), 
an upper and lower limit of carbon abundances (dashed blue line).
}
\label{comp}
\end{figure*}

\subsubsection{J012512+070319}

The quality of the BOSS spectrum of this target is the highest in the sample. The initially derived set of 
parameters appears to be reliable; the much higher S/N ratio OSIRIS spectra lead to an effective 
temperature that differs by only 140\,K. The slightly lower temperature we find decreases the derived 
metallicity by 0.1\,dex. The final values for the parameters and metallicities of this object are 
$\rm T_{eff}=6312\pm 78$\,K,  
$\logg=5.0\pm0.5$, and  [Fe/H]$=-3.3\pm0.2$. The derived carbon abundance is [C/Fe]$=1.2\pm0.6$.

\subsubsection{J015131+163944}

This object was observed with ISIS on the WHT and analyzed by \citet{agu17I}. 
The set of derived parameters by these authors is $\rm T_{eff}=6025\pm 77$\,K,  $\logg=4.8\pm0.6$,
[Fe/H]$=-3.8\pm0.2$ and [C/Fe]$=1.3\pm0.4$, from a spectrum with very similar resolving power 
(R$\sim2400$) and slightly lower S/N ratio than the OSIRIS spectrum presented here.
The values given in Table \ref{AnalysisResults} are essentially the same:
$\rm T_{eff}=6036\pm 75$\,K,  $\logg=5.0\pm0.5$, [Fe/H]$=-3.8\pm0.2$ and [C/Fe]$=1.1\pm0.5$.

\subsubsection{J041800+062308}

The SEGUE spectrum of this object has a fairly poor quality, and, as a result, the derived effective temperature, 
$\rm T_{eff}=5579$\,K, is not very reliable. The set of parameters from the OSIRIS analysis
($\rm T_{\rm eff}=6247\pm 78$\,K, $\logg=5.0\pm0.5$, [Fe/H]$=-3.4\pm0.2$) are more robust, while the
G-band is close to the detection limit ([C/Fe]$=0.7\pm0.7$).

\subsubsection{J094708+461010}\label{j0947}

The S/N of this BOSS spectrum is the lowest in the sample, and for this reason we consider the derived
atmospheric parameters in Table \ref{basic} unreliable. However, both the effective temperature and the 
gravity derived from the OSIRIS spectrum are compatible with those from the BOSS observations. Accepting 
the FERRE derived values, we arrive at 
$\rm T_{eff}=5858\pm 73$\,K,  $\logg=5.0\pm0.5$, [Fe/H]$=-4.1\pm0.2$, and [C/Fe]$=1.0\pm0.4$. Interestingly enough, the 
Yoon-Beers diagram described in \citet{yoo16} points out that there are two different “families” of
CEMP-no stars, which they designate as Group II and Group III. J094708+461010 is a Group II with [Fe/H]$=-4.1$ and A(C)=5.3.

\subsubsection{J173403+644632}

The spectrum of this target shows the clearest and unresolved calcium ISM contribution, 
and the strongest G-band absorption, making this faint object very interesting candidate for a detailed study.
Since the \ion{Ca}{II} H-K spectral region (the one containing more information) is affected by CH lines, 
we perform the analysis with FERRE as follows:

\begin{itemize}
 
\item We derived simultaneously the effective temperature ($\rm T_{eff}=6183$\,K (see Fig. \ref{comp}, 
upper panel), the carbon abundance
([C/Fe]$=3.1$: see Fig. \ref{comp}, lower panel), and a surface gravity of $\logg=5.0$; 

\item A Gaussian profile for the  calcium ISM contribution is adopted. 
A grid of varying absorption coefficients ($\rm \epsilon$) from 5\% to
30\% in step sizes of 1\% and relative velocities from 180\,km s$^{-1}$ to 220\,km s$^{-1}$ 
in steps of 5\,km s$^{-1}$ were subtracted, and thus the resulting spectrum was reanalyzed with FERRE.  

\item The minimum $\chi^{2}$ is obtained for the point $\rm \epsilon=$0.29 and $V=195$\,km s$^{-1}$ which 
reproduces best both the \ion{Ca}{II} and the $\rm H_{\gamma}$ lines (see Fig. \ref{comp}, upper panel).

\end{itemize}

Thus, the final set of parameters are
$\rm T_{eff}=6183\pm 78$\,K,  $\logg=5.0\pm0.5$, [Fe/H]$=-4.3\pm0.2$, and [C/Fe]$=3.1\pm0.2$. Again, regarding the diagram mentioned in Section\ref{j0947}, J173403+644632 with [Fe/H]$=-4.3$ and A(C)$=7.2$ would clearly be in the Group III This 
target is the only ultra metal-poor star known at $V>$19.
 Finally, the fact that the derived gravities are always on the limit of the grid is compatible with the relatively high error bars adopted. However, the influence on the derived metallicity and carbon abundance is small and  already considered in error bars through other sources of error \citep{agu17I}.

\section{Conclusions}\label{discuss}

The multiple techniques use to identify and confirm metal-poor stars
have been summarized by \citet{bee05},\citet{fre15rev}, 
and references therein, and include high proper-motion surveys, Schmidt objective-prism surveys, 
photometric surveys, and spectroscopic surveys.
The most iron-poor star known, SM 0313$-$6708 with [Fe/H]$<-7.3$,
was discovered from a photometric survey, the Skymapper Southern Sky Survey \citep{kel14}.
The most metal-poor star known, J1029+1729, was in turn  recognized in the 
SDSS spectroscopic survey by \citet{caff12I}.
Obviously, the observational efforts and resources employed demand less telescope timewhen using photometry to identify metal-poor stars  \citep{sta17}, 
but the only path to confirm and derive reliable metallicities involves spectroscopic techniques.

Usually a low-resolution spectroscopic analysis is required to securely identify 
metal-poor candidates, and the subsequent and significantly more expensive high-resolution 
analysis confirms the metallicity and provides multiple chemical abundances.
Avoiding false positives (incorrect identifications) optimizes the invested time in 6-10\,m class telescopes.
In this paper we use an intermediate step \citep[see][]{bee85} to increase this ratio. 
Using medium-resolution spectroscopy we are able to derive reliable metallicities and 
sometimes carbon abundances, even in the [Fe/H]$<-4.0$ regime.

\begin{figure}
\begin{center}
{\includegraphics[width=90 mm, angle=180]{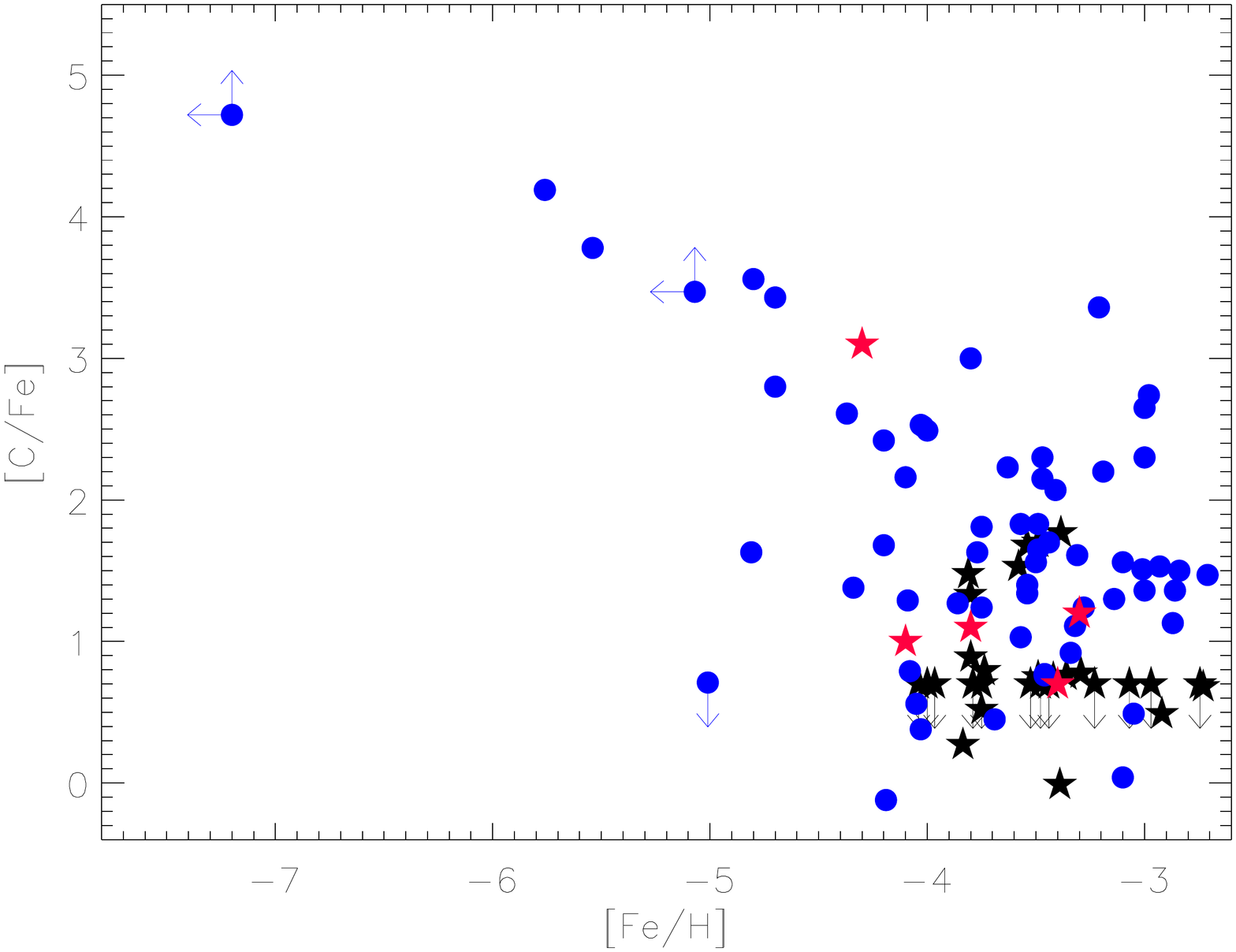}}
\end{center}
\caption{Carbon over iron versus metallicities of stars presented in this work (red filled stars) 
and those from \citet{agu17I} (black filled stars) analyzed with 
the improved methodology, 
together with stars from the literature  
 \citep{siv06,yong13II,fre05,fre06,caff14,alle15} 
represented by blue filled circles.}
\label{carbon}
\end{figure}

We have identified a number of very faint metal-poor stars, 
and offer to the community a sizable sample of faint ultra metal-poor stars to observe 
with the next generation of 30--40\,m telescopes.
Figure \ref{carbon} shows the carbon abundances derived from OSIRIS spectra for the targets 
studied in this work, together with those from \citet{agu17I}, and other CEMP stars 
from the literature. The chance to carry out a follow-up program for observing stars with V>19 significantly improves  our success ratio of identifying ultra metal-poor stars. The more carbon-enhanced metal-poor stars are those that are below the [Fe/H]=-4.0 regime.

In addition, we demonstrate the advantages of our methodology when the calcium ISM contribution is not resolved at medium resolution which is the vast majority. However, the most interesting star of this work, J173403+644632, is an excellent candidate to observe at higher resolution spectroscopy and to derive its chemical composition in detail, giving important constraints to the early Universe. Finally,  we have observed and analyzed, with highly reliable results, well-known metal-poor stars (G64-12 and J1029+1729), together with those presented in \citet{agu17I}, and this allows us to trust our methodology when looking for ultra metal-poor stars.

\begin{acknowledgements}
D.A. acknowledges the Spanish Ministry of Economy and Competitiveness (MINECO) for the financial support
received in the form of a Severo-Ochoa PhD fellowship, within the Severo-Ochoa International PhD Program.
D.A., C.A.P., J.I.G.H., and R.R. also acknowledge the Spanish ministry project MINECO AYA2014-56359-P. 
J.I.G.H. acknowledges financial support from the Spanish Ministry of Economy and Competitiveness (MINECO)
under the 2013 Ram\'on y Cajal program MINECO RYC-2013-14875. 

This paper is based on observations made with the Gran Telescopio de Canarias,
operated by the GRANTECAN team at the Observatorio del Roque de los Muchachos, La Palma,
Spain, of the Instituto de Astrof{\'i}sica de Canarias. We thank the GRANTECAN 
staff members in general and Antonio Cabrera Lavers and Daniel Reverte Pay\'a in particular 
for their efficiency during the program preparation.  \\
\end{acknowledgements}

%
%

\bibliography{biblio}

\end{document}